\documentclass[aps,prl,twocolumn,showpacs]{revtex4}
\usepackage{epsfig}
\usepackage{graphicx}
\usepackage{xcolor}

\begin{document}

\title{Note on the reflectance of mirrors exposed to a strontium beam}

\author{J. Huckans$^{3}$, W. Dubosclard$^{1,2}$, E. Mar\'echal$^{2,1}$, O. Gorceix$^{1,2}$, B. Laburthe-Tolra$^{2,1}$, M. Robert-de-Saint-Vincent$^{2,1}$}

\affiliation{$^1$ Universit\'e Paris 13, Sorbonne Paris Cit\'e, Laboratoire de Physique des Lasers, F-93430
Villetaneuse, France \\$^2$ CNRS, UMR 7538, LPL, F-93430 Villetaneuse, France \\$^3$ Department of Physics and Engineering, Bloomsburg University, Bloomsburg, Pennsylvania}

\DeclareGraphicsExtensions{.eps,.EPS,.pdf}

\newpage
\begin{abstract}

\textit{We here share a note on reflectivity tests on mirrors exposed to a strontium atomic beam. Unfortunately, insufficiently high vacuum conditions prevent our results from being conclusive for the intended application of Zeeman slowing in ultra-high vacuum. We nevertheless hope that this note may be useful to teams realizing similar tests.}

~\\
The high chemical reactivity of strontium, which can opacify a viewport exposed to a strontium atomic source, is a concern for some atomic physics experiments where it is sometimes necessary to send a laser beam counter-propagating relative to the atomic beam. While a number of experiments use heated sapphire windows to reduce strontium deposition and increase the viewport lifetime, here we study another possibility, consisting of sending the laser beam into the atomic flux by reflecting it off a mirror at 45$^{\circ}$ exposed to the strontium flux. 
We present our attempt to find a substrate that can be exposed to strontium and maintain high reflectivity.  
We first present the formation of a strontium metallic mirror under high flux ($> 10^{13}$ at/s/cm$^2$) on a sapphire substrate, and measure its reflectivity at 45$^\circ$ to be 0.65 (S) and 0.51 (P). On two other substrates, initially reflective metallic mirrors, we show for slightly lower fluxes (i.e., a factor of 3) that some reaction - most probably oxidation - is able to prevent the formation of the metallic layer even in high vacuum conditions. Instead, we observe the growth of a dielectric transparent medium. Despite the continuous deposition of strontium, the back surface reflectivity continues to dominate. 
We show the unusual evolution of reflectivity on these substrates, and emphasize two observations: i) a sharp threshold in the strontium flux separating transparent material growth from lossy material growth; ii) strontium's highly efficient capture of oxygen, even from rarefied sources: here mostly the residual high vacuum pressure (10$^{-7}$mbar full pressure) and possibly a protective SiO$_2$ surface on one of the substrates.
\end{abstract}

\date{\today}
\maketitle

\section{Introduction}

Part of the complexity of atomic physics experiments lies with their atomic sources. For atomic ovens and 2d-MOTs, considerations involving vapor pressures, chemical reactivities, and even high-temperature metallurgy may come into play \cite{bechtoldt1957}. A common occurrence is that the atomic flux exiting the source coats vacuum windows so that they become opaque. For several alkali species, laser induced atom desorption solves this problem and may even turn it into an advantage by the creation of an ambient temperature, low-residual-gas pressure source \cite{Gozzini,Klempt}. For some other species, like chromium and erbium, the issue of the viewport facing the source, used to insert a Zeeman slowing beam, is circumvented by the use of a 45$^{\circ}$ mirror to insert the beam from the side. This mirror is slowly coated by the atomic beam \cite{Bell,Chicireanu,Frisch2014} and maintains a reasonable long term reflectance.  For strontium, which exhibits strong reactivity even with standard BK7 glass, currently the most widely used solution is to prevent the formation of a coating by a) choosing  a viewport material with low chemical reactivity, and b) increasing the viewport's temperature (at least 150$^{\circ}$C for a sapphire viewport) \cite{Stellmer2012,Bongs2015}, thus compensating adsorption with desorption. However, a failure to maintain a sufficiently high temperature may result in the formation of a thin film that is not removable in a reasonable amount of time by returning to high temperature. This results in major inconveniences: poor vacuum, a surface without an anti-reflection coating, typically additional valves and vacuum elements in order to replace the damaged viewport. Therefore, other complex strategies have been devised, such as diverting the atomic beam between the oven and the high vacuum trapping region \cite{yang2015, Nosske2017}.

In this work, our aim was to study the feasibility of installing a mirror under vacuum, with a 45$^\circ$ orientation that may be used to redirect the Zeeman slower laser beam along the strontium beam propagation axis. We wished to answer the following basic question : how much and at what rate will the reflectance of this mirror change when subjected to the strontium beam?  The question of the reactivity with the surface complicates the matter. As a benchmark, we first deposited Sr on a pure polished sapphire substrate, and we observed the building of a reflective surface after about 100 nm of Sr deposition, i.e. 0.26 g/m$^2$ producing a mirror with a moderate reflectance, 65\% for S polarization and 51\% for P polarization. We observed also that under our vacuum conditions (about $1\times\,10^{-7}$\,mbar), this mirror was unstable within a few days. We then monitored the reflectance of two mirrors with good initial reflectances ($>90\,\%$) at 461 nm, both subjected to a continuous deposition of strontium: an SiO$_2$-protected aluminum mirror, and a silver mirror, unprotected. We hoped to find a mirror compatible with long term strontium beam exposure, as well as to quantify the performance and the lifetime of the exposed mirrors. For the deposition on the two mirrors, we originally surmised that after a deposition thickness of the order of 100 nm, we would observe the continuous change of the reflectance of the bare mirror to a value given by metallic strontium. Instead, we found that for the SiO$_2$-protected Al mirror, the initially high reflectance of Al was maintained for a Sr deposition thickness up to 3.2 g/m$^2$. For the unprotected Ag mirror, we observed a similar preservation of the reflectance, but for a thinner deposition. 

Ultimately, we observed that strontium had in the end a destructive effect on all three substrates : after a week of exposure, corresponding to a deposited thickness of several micrometers, the Al and Ag mirrors' reflectances fell below 20\%. Under our experimental conditions, the deposition flux of about 10$^{13}$at/s/cm$^2$ is typically 10$^3$ time higher than the standard flux for strontium cold atom experiments. In the best case of the SiO$_2$ protected Al mirror, the one week timescale (to reach 3.2 g/m$^2$ thickness) resulting in failure of the mirror for our experimental condition would correspond to several years at standard flux conditions for a cold atom experiment. We associate the surprisingly long lifetime of the mirror to oxygen intake from the residual water pressure, which probably dominated that from the SiO$_2$ surface. It would be interesting to investigate whether the SiO$_2$ surface, under ultra-high vacuum conditions, would provide sufficient oxygen to have a similar effect on small strontium depositions and thus maintain for an extended time the underlying higher reflectivity of commercial protected metallic mirrors.

\section{experiment set-up and flux considerations}

\begin{figure}[htbp]
 \centering 
\includegraphics[width=0.7\linewidth]{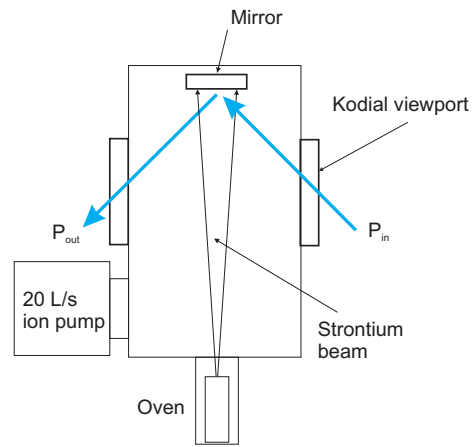}
\caption{Experimental set-up. A 1" diameter mirror is located 30 cm from the oven output, on the beam axis. The reflectance evolution at 461 nm is obtained by continuously monitoring  the laser powers P$_{out}$ and P$_{in}$ after and before the mirror on two separate photo diodes (not shown), and taking the ratio of these two values. The input polarization (either S or P) can be changed with a wave plate.}
\label{Fig1}
\end{figure}

A schematic of the setup is presented in Fig.\,\ref{Fig1}. The strontium reservoir is a stainless steel tube (12 mm external diameter, 40 mm length) initially filled with 2 g of strontium pieces with 2N (99\%) purity. The reservoir output aperture (4 mm diameter, 15 mm length) is filled with 44 stainless steel micro-tubes (r = 190 $\mu$m internal radius, 15 mm length). These micro-tubes create an effusive, directional beam, and also considerably increase the lifetime of the oven before refilling \cite{Schioppo}. The reservoir is inserted inside a DNCF16 nipple connected to the vacuum chamber. This tube is externally heated and the temperature is monitored using a thermocouple. The vacuum is maintained by a 20 l/s ion pump. During the deposition, the pressure at the mirror position is of the order of $1\times 10^{-7}$ mbar.

The tested substrate is located at a distance L\,=\,30 cm from the oven output. Two Kodial uncoated DNCF40 viewports are used to send a 461 nm laser beam  (nominal 500 $\mu$m waist, 1 mW) onto the exposed mirror with an incidence angle of 45$^\circ$. The laser frequency is kept away from the atomic resonance. The ratio of the output to the input beam power, measured continuously on two separate photo-diodes, is used to monitor the evolution of the reflectance as a function of time, for S or P input polarizations.

A simple estimate of the strontium beam intensity at the beam center (atomic flux per  unit surface)  at a distance $L$ from the oven output is given by $I_{theo}=N_{\mu tubes}\times \frac{\pi r^2\bar{v}}{4\pi }\frac{n}{L^2}$ \cite{Ramsey} where $r$ is the micro-tube radius, $n=P/k_BT$  and $\bar{v}=\sqrt{8k_BT/\pi M}$ respectively the strontium density inside the reservoir and the mean velocity along the beam at a temperature $T$. $P$ is the strontium vapor pressure inside the oven and can be calculated using the following formula : $P(Pa)=10^{14.232-\frac{8572}{T(K)}-1.1962 \log_{10} T(K)}$ \cite{Schioppo}. This simple analytical expression was compared with a direct measurement using the linear absorption of a laser beam tuned at resonance with the broad line $5s^2$ $^1S_0$ $\leftrightarrow$ $5s5p$ $^1P_1$ strontium transition at 460.862  nm  ($\Gamma$=\,32 MHz) \cite{NoteAbs}.
At the temperatures between 520$^\circ$C and 560$^\circ$C used for strontium deposition in this work, atomic beam intensities $I$ were tested from $5.3\times 10^{12}$ at/s/cm$^2$ to $1.7\times 10^{13}$ at/s/cm$^2$ \cite{NoteBis}.
At the center of the atomic beam, the growth rate of  a metallic strontium layer would be given by $K=I\times M/\rho$ where $\rho=2.64$ g/cm$^3$ is the density of metallic strontium, thus spanning from 11 to 34 nm/hour.

\section{sapphire substrate as a benchmark}

Strontium is a highly reactive species, and it oxidizes quickly when in contact with the ambient atmosphere. The reflectivity of strontium is therefore poorly documented. We decided to deposit strontium on a substrate with low reactivity to characterize its optical properties. We chose an optical quality sapphire surface (Al$_2$O$_3$).  Sapphire is often used as the entrance window for Zeeman slowing beams for strontium, since strontium does not chemically react with it and elevated temperatures (about 150$^{\circ}$C) are usually sufficient to prevent opaque layer buildup.  Left at room temperature, it would appear to be a good prospect for depositing a metallic, reflective strontium layer.

\begin{figure}[htbp]
\includegraphics[width=\linewidth]{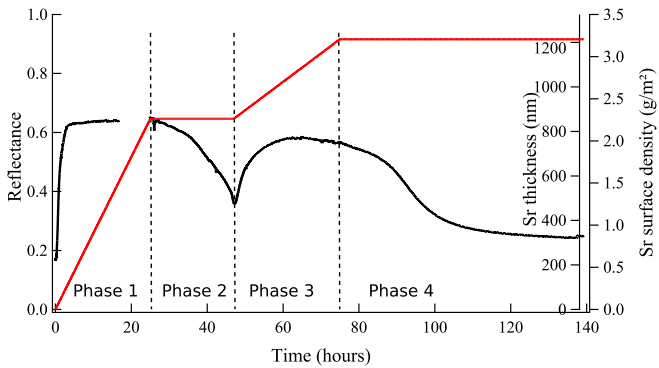}
\caption{Black and left axis: Reflectance of 461 nm S-polarized laser light during deposition of strontium on sapphire as a function of deposition time.  Phase 1 : constant initial flux of strontium $1.7\,10^{13}\,$at/s/cm$^2$.  Phase 2: cessation of strontium flux.  Phase 3: constant strontium flux, restored to $6.4\,10^{12}\,$at/s/cm$^2$. Phase 4: cessation of strontium flux. Red and right axes : estimated deposition. The strontium surface density is estimated from the flux, while the film thickness, on the first right-axis scale, furthermore assumes that the film grows as a metallic strontium layer.}
\label{Fig2}
\end{figure}

Measurements were made over a period of approximately six days. The strontium beam intensity  was initially raised to $(1.7\pm0.5)\,10^{13}\,$at/s/cm$^2$ corresponding to a  deposition rate of order 34 nm/hour at the center of the beam.  As shown in Fig.\,\ref{Fig2} (phase 1), although the bare substrate (a dielectric) is initially a very poor 461 nm reflector with a 16\% reflectance per face for S polarization, the deposition of strontium quickly leads (after $\sim$3 hours) to a reflectance of 65\% for S polarization (but only of 51\% for P polarization).  We observe therefore that a mirror is formed after a deposited thickness of $\sim$100 nm. This illustrates the good chemical inertia of sapphire that does not seem to react with strontium, such that a metallic layer is created. The measured reflectances of strontium of 65\%  and 51\% at 45$^\circ$  for S and P polarization respectively  are consistent with what is expected from the published value of 60 \% at normal incidence at 461 nm \cite{IndexStrontium}. (Indeed, for a standard metal, the S reflectance increases monotonically with angle from the value at normal incidence to almost 100 \% at 90$^\circ$ degree incidence, and the P reflectance shows a Brewster like minimum).

We found also that under our experimental conditions, the deposited strontium mirror was not stable as shown in Fig.\,\ref{Fig2}\,: when the flux was turned off and the deposition interrupted (phase 2), the reflectance slowly diminished. After restoring the atomic flux (phase 3), to a lower value of $(6.4\pm1.2)\,10^{12}\,$at/s/cm$^2$, reflectance increased again but not as high as initially. Again (phase 4), the oven was turned off and the reflectance slowly decreased within about one day.  
We think that the mirror degradation might have been caused by residual molecules in our vacuum (of order 10$^{-8}$\,mbar during phases 2 and 4) reacting with the thin film.  It is possible that the strontium reflectance could have been maintained for a substantially longer time if the vacuum were in the range of typical cold atom experiments, $10^{-11}$ mbar. Nevertheless, the formation of an in-situ mirror using a bare sapphire substrate is difficult to implement in a standard cold atom experiment, as a high flux should be used initially. This deposited sample is used as a benchmark for the following studies where we have installed an initially good mirror at 461 nm.

\section{SiO$_2$-protected aluminium mirror}

\begin{figure}[htbp]
\includegraphics[width=\linewidth]{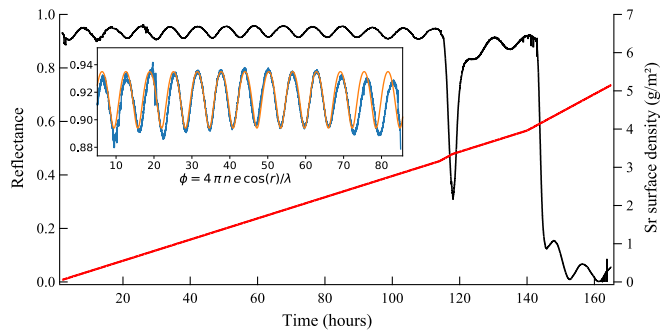}
\caption{Black: Reflectance of P-polarized 461 nm laser light during deposition of strontium on SiO$_2$-protected Al as a function of deposition time - first four days of reflectance measurements. 
Red : deposited strontium surface density, estimated from the flux. Inset: comparison between data (blue) and a multiple-layer thin film model, during the first 115 hours, based on transfer matrices (see text). From the fringe contrast we derive the index of the optical material, while the periodicity further provides an estimate of the thickness growth rate, independent of any assumption about the material. }
\label{Fig3}
\end{figure}

We monitored the reflectance of an Al mirror protected by 100 nm of SiO$_2$ \cite{MirrorThorlabs} exposed to the strontium beam. Measurements were made over a period of approximately eight days, for P input polarization of the laser, at 461 nm and under 45$^{\circ}$ incidence. The initial flux is estimated to $(5.3\pm1.6)10^{12}\,$at/s/cm$^2$, about a factor of 3 below the flux on sapphire during phase 1, and within error range from the flux on sapphire during phase 3. As shown in Fig.\,\ref{Fig3}, starting from a bare, strontium-free reflective surface, we observed high average light reflectance ($\sim$0.89) for more than 100 hours, which sinusoidally varied with 0.02 amplitude with a period of 8.5 hours. The average reflectance is equal to the expected value of Al at this wavelength for P polarization. We emphasize that our results are surprising : one single oscillation of 8.5 hours  corresponds to a deposition of about 0.24 g/m$^2$, which would build 90 nm of metallic Sr. According to our observations on sapphire, this should suffice to create a mirror. Were such a mirror built, the reflectance would be lower. We suspect that strontium has instead formed a transparent dielectric material by oxidation and we interpret the observed small amplitude oscillations as interference fringes as the dielectric thickness grows. 

Using the transfer matrix formulation for the description of multiple thin films \cite{matrices}, we model the observed oscillations by the growth of an unknown dielectric, on top of the 100\,nm layer of SiO$_2$, itself on aluminum. First, the phase of the observed oscillations is extracted by a simple sinusoidal fit to the data: the time period $T$ is matched to a $2 \pi$ phase increase. The periodicity is not perfect, highlighting some irregularity in the growth dynamics. Then, we evaluate with the thin film model the reflectivity as a function of the phase $\phi = 4 \pi n e \cos(r) / \lambda$, where $e$ is the thickness of the material, $n$ its refractive index, and $r$ the angle of refraction. Here, the reflectivity of aluminum is set from that at the refracted angle inside the SiO$_2$ material ($28\,^\circ$): $R_{Alu}^P = 0.91$. Only the optical index $n$, controlling the fringe contrast in the calculation, and a small fine-tuning of the absolute reflectance measurement are used to fit the model on the data (see Fig.\,\ref{Fig3}\,inset). We thus obtain a measurement of $n = 1.65 \pm 0.07$ that relies neither on any assumption on the nature of the material nor on any deposition rate estimate. Furthermore, we infer the thickness periodicity from the model $\delta e = \lambda / 2 n \cos(r)$, which compared to the observed time periodicity provides a growth rate of the material $\delta e/T \simeq 18$\, nm/hours, without assumption on its nature. 

We now briefly speculate on the mechanism that leads to the formation of a transparent layer. Strontium may have been partially oxidized, by the residual water vapor pressure or by reaction with SiO$_2$. A full oxidation would produce SrO, which has an optical index of 1.86. Assuming the order of magnitude of density $\rho$ and molecular mass $m$ from those of Sr metal and of SrO, the growth rate deduced from the thin film model is compatible with our initial estimate of the atomix flux:  $I \sim \frac{\delta e}{T} \frac{\rho}{m}$.
The 100\,nm SiO$_2$ layer may provide 0.14 g/m$^2$ of oxygen, enough atoms to fully oxidize about 25$\%$ of the 3.2\,g/m$^2$ deposit for which transparency is consistently observed.  
On the other hand, the ambient pressure could also provide a similar or higher quantity of oxygen atoms, by the continuous deposition of water molecules on the surface.
The full pressure on this substrate is of order $10^{-7}$ mbar, and typically higher than during the deposition on sapphire. The partial water pressure is not measured. Assuming for example 10$\%$ of water pressure \cite{Rommel}, and assuming a 100$\%$ capture efficiency of incoming water molecules, the oxygen deposit  in the first $t = 115$ hours, would be $\rho^{surf}_O = \frac{1}{4}\frac{P}{k_B T}\sqrt{ \frac{8 k_B T}{\pi m_{\rm H_2O}}} m_{\rm O} t \approx 0.4~$g/m$^2$ of oxygen (here, P = 10$^{-8}$ mbar is the partial water pressure, T = 293\,K is the chamber ambient temperature, $m_{\rm H_2O}$ and $ m_{\rm O}$ are the masses of the water molecule and the oxygen atom).

More complex phenomena were observed when changing the flux. After 115 hours, we suddenly increased the atomic flux by roughly a factor of two by increasing the oven temperature from 540$^\circ$C to 560$^\circ$C \cite{NoteBis}.  This had the effect of rapidly reducing the mirror reflectance by roughly a factor of three. After approximately six hours at this higher flux, we suddenly reduced the oven temperature to the initial value, thus restoring the initial flux. The reflectance quickly increased and approached the original value, while the oscillations continued. We speculate that at higher flux, the oxygen intake was not fast enough to form the dielectric material, leading to the formation of a lossy (absorbing or diffusive) material. Restoring the initial flux, allowed for the formation of a transparent dielectric so that a good reflectance was recovered. After nearly one day at the original flux, we again increased the flux of the oven by the same amount. The reflectance suddenly collapsed to about 0.2 average. High reflectance was never regained despite various strategies.  As shown in Fig. \ref{Fig3}, the estimated strontium deposit at the end of this exposure was about 5.2 g/m$^2$.

\begin{figure}[htbp]
 \centering
\fbox{\includegraphics[width=0.9\linewidth]{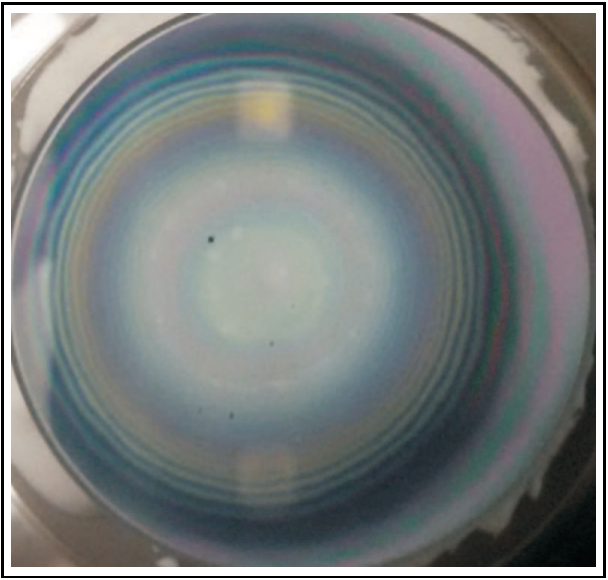}}
\caption{Photograph of the exposed SiO$_2$ protected aluminium mirror, removed from vacuum after deposition, displaying equal-thickness interference fringes.}
\label{Fig4}
\end{figure}

We opened the vacuum chamber to inspect the surface of the aluminum mirror.  As seen in Fig.\,\ref{Fig4}, complete oxidation rapidly affected the substrate, leading to the formation of a white diffusive semi-transparent material. Although the central region appears white in Fig.\,\ref{Fig4}, it was originally black - thus not reflective - and turned from black to white in approximately 15 seconds upon exposure to air. Fringes are interpreted as equal thickness contours in the deposit, thickest at the center and vanishing at the edge. The fringes result from reflections off the outer surface of the strontium film (oxidized by air in the figure) and of the SiO$_2$-protected aluminum/strontium interface below it. The roughly 20 visible fringes observed in Fig.\,\ref{Fig4} are also consistent with the measured 20 light intensity oscillations shown in Fig.\,\ref{Fig3}. Following our visual inspection, we scraped the majority of the film off the mirror and measured a yield of $m=4\times10^{-4}$ g (one order of magnitude above the initial mass of SiO$_2$ on the scraped surface).  The strontium was deposited in about 160 hours. Assuming a molar mass of the weighted molecules of 104 g/mol (SrO) and Gaussian deposition profile with 7 mm 1/e$^2$ half width we obtain a second estimate of the average strontium atomic flux, $I^{(2)} = 5.2\,10^{12}$\,at/s/cm$^2$, in agreement with the first one.

We conclude that, in our vacuum conditions, for the deposition of strontium on SiO$_2$-protected Al mirrors the reflectance remains close to that of the Al surface for deposition of 3.2 g/m$^2$ of strontium. We observe only 0.04 contrast oscillations in the reflectivity for $P$ polarization. According to the thin film model, for $S$ polarization the contrast would be larger, with reflectance oscillating between 0.87 and 0.97; thus, for polarization-sensitive applications, the polarization would need periodic adjustment, but power losses remain acceptable. Higher flux and/or thicker deposition leads to the degradation of the mirror.

\section{Unprotected silver mirror}

\begin{figure}[htbp]
\includegraphics[width=\linewidth]{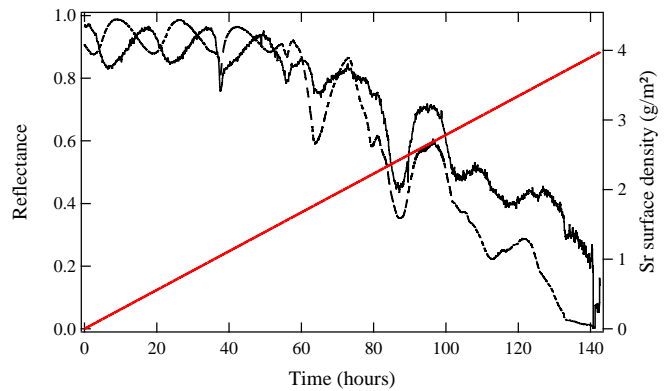}
\caption{Black : Reflectance of 461 nm laser light during deposition of strontium on silver as a function of deposition time.  Solid line: S polarization.  Dashed line: P polarization. Red : deposited strontium surface density.} 
\label{Fig5}
\end{figure}

As an alternative, we tested deposition on a substrate composed of bare silver. It was fabricated at our institute LPL by vapor deposition on a glass substrate and moved to our experimental chamber under nitrogen. A short exposure to atmospheric oxygen for several tens of seconds still occurred. Reflectance measurements were made over a period of approximately six days with vacuum pressures in the low $10^{-7}$ mbar range. For this measurement, we implemented a more complete polarization setup allowing us to monitor simultaneously the reflectance for P and S input polarization. Peak atomic flux at the surface of the mirror is approximately as for deposition on aluminum, $I \simeq (5.3\pm1.6)10^{12}\,$at/s/cm$^2$, constant throughout the entire experiment.
An interesting feature of these measurements (see Fig.\,\ref{Fig5}) was the observation of out-of-phase reflectance oscillations between S and P polarizations, at early time, without significant change of the average reflectance. After two days, an increasingly strong in-phase oscillation took over, corresponding to actual power loss, coincident with a steady loss of the average reflectance. 
The early oscillations may reveal birefringence in the deposit, while the later oscillations may be interference fringes, as on the SiO$_2$-protected substrate but in a lossy medium. The distinction between the two regimes is only speculation. After $\sim$ 1 g/m$^2$ deposit, oxygen from the partial water pressure or from the silver surface may be running out;
the deposit then builds up as a material with unsuitable optical properties  (absorbing or diffusive). After six days and deposition of 4 g/m$^2$, the average reflectance was less than 0.2. High reflectance was never regained and the measurements were terminated. In this case as well, long term strontium deposition on a silver mirror led to the destruction of the mirror.

\section{Conclusion}

Our study highlights that the very high reactivity of strontium plays a significant role during deposition even in high vacuum conditions. On the sapphire substrate, after measuring the reflectivity of metallic strontium, we observed that a pressure in the $10^{-8}\,$mbar range (the partial water pressure being unknown) is sufficient to destroy an almost $\mu$m thick mirror in a matter of days. Furthermore, in some conditions strontium reactivity may play a positive role preserving a high reflectance. Starting from a good reflecting mirror,  here a commercial Al mirror protected by a thin layer of SiO$_2$, for vacuum pressures in the 10$^{-7}$\,mbar range, a reaction involving strontium, probably oxidation, built a transparent layer that does not occlude the back surface aluminum reflectivity. This leads to a fifteen fold extension of the expected lifetime of the mirror: we demonstrated a preservation of the aluminium reflectance for up to 3.2 g/m$^2$ of strontium, enough to build a metallic Sr layer of thickness 1.2 $\mu$m. This preservation of the mirror reflectance could be useful to insert a light beam at a right angle relative to the atomic source beam. 

The conditions of this preservation are tied to the strontium flux, to the substrate, and to the ambient pressure. On the SiO$_2$ protected aluminum substrate, a sharp cutoff in flux separates transparent material growth from lossy material growth. The simplest explanation for our observations is a critical ratio between strontium deposition rate and ambient water collisions on the surface. 

Because the atomic beam intensities in our experiments are high compared to some strontium experiments, due to the high temperature and short distance (30\,cm) between oven and mirror, cancellation of the strontium reflectivity may be relevant to experiments in better vacuum conditions - should the reduction in atomic beam intensity equate or exceed the reduction in partial pressures of water and oxygen. Our beam intensity during the transparent deposition is for example of order 600 times that of the $^{88}$Sr compact atomic clock \cite{Poli}. Furthermore, the quantity of oxygen available in the SiO$_2$ layer also being significant, it would be interesting to see in true ultra-high vacuum conditions, with water partial pressure below 10$^{-11}$ mbar, whether similar cancellations of the strontium reflectivity may be observed, and for which thicknesses and deposition rates.

We acknowledge T. Billeton for the fabrication of the silver mirror, A. Kaladjian, F. Wiotte and and H. Mouhamad for technical support, and D. Bloch for discussion of these results. We  acknowledge  funding  support  from  the  Agence  Nationale de la Recherche (ANR Tremplin-ERC Highspin ANR-16-TERC-0015-01 and Labex First-TF ANR-10-LABX-48-01), the Conseil Régional d'Ile-de-France (Domaine d'Intérêt Majeur Nano'K, Institut Francilien des Atomes Froids), and Bloomsburg University Foundation.


\begin{thebibliography}{199}
\bibitem{NoteAbs} We compare the absorption measurement with a calculation that takes into account the velocity distribution for our oven geometry at temperature T.
\bibitem{NoteBis} During deposition on the protected aluminium and on the silver substrate, a cold aperture was partially obstructing the beam, reducing the flux by a factor 3.2. It was removed for the Sr deposition on Al$_2$O$_3$.
\end{thebibliography}
\end{document}